\documentclass[12pt]{article}\usepackage[hyperfootnotes=false]{hyperref}
\usepackage{epsfig}
\usepackage{amsmath}
\usepackage{amssymb}
\usepackage{graphicx}
\newlength{\abstractwidth}
\renewcommand{\thefootnote}{\fnsymbol{footnote}}
\renewcommand{\thanks}[1]{\footnote{#1}} 
\newcommand{\starttext}{
\setcounter{footnote}{0}
\renewcommand{\thefootnote}{\arabic{footnote}}}
\renewcommand{\theequation}{\thesection.\arabic{equation}}
\newcommand{\be}{\begin{equation}}
\newcommand{\bea}{\begin{eqnarray}}
\newcommand{\eea}{\end{eqnarray}}
\newcommand{\beq}{\begin{equation}}
\newcommand{\ee}{\end{equation}}
\newcommand{\eeq}{\end{equation}}

\def\ba{\begin{eqnarray}}
\def\ea{\end{eqnarray}}

\def\12{{1 \over 2}}
\def\eq{&=&}

\def\ra{\rangle}

\def\simleq{\; \raise0.3ex\hbox{$<$\kern-0.75em
\raise-1.1ex\hbox{$\sim$}}\; }
\def\simgeq{\; \raise0.3ex\hbox{$>$\kern-0.75em
\raise-1.1ex\hbox{$\sim$}}\; }

\def\O2{\Omega_2}

\def\bi{\begin{itemize}}
\def\ei{\end{itemize}}

\def\sc{\setcounter{equation}{0}}

\def\W{$\Omega$}
\def\W'{$\Omega$}

\def\V{\Omega}
\def\V'{\Omega}

\def\O{${\cal{O}}$}

\def\c{{\cal{C}}}

\def\t{{\cal{T}}}

\def\t{\tau}

\makeatletter
\g@addto@macro\normalsize{%
  \setlength\abovedisplayskip{10pt}
  \setlength\belowdisplayskip{20pt}
  \setlength\abovedisplayshortskip{10pt}
  \setlength\belowdisplayshortskip{20pt}
}
\makeatother

\usepackage{color}

\begin{document}
\renewcommand{\theequation}{\thesection.\arabic{equation}}
\begin{titlepage}
\rightline{}
\bigskip
\bigskip\bigskip\bigskip\bigskip
\bigskip
\centerline{\Large \bf { Complexity and Shock Wave Geometries}}
\bigskip

\bigskip
\begin{center}
\bf Douglas Stanford and Leonard Susskind  \rm

\bigskip

\bigskip

Stanford Institute for Theoretical Physics and Department of Physics, \\
Stanford University,
Stanford, CA 94305-4060, USA \\
\bigskip
\bigskip
\vspace{2cm}
\end{center}
\bigskip\bigskip
\bigskip\bigskip
\begin{abstract}

In this paper we refine a conjecture relating the time-dependent size of an Einstein-Rosen bridge to the computational complexity of the dual quantum state. Our refinement states that the complexity is proportional to the spatial volume of the ERB. More precisely, up to an ambiguous numerical coefficient, we propose that the complexity is the regularized volume of the largest codimension one surface crossing the bridge, divided by $G_N l_{AdS}$. We test this conjecture against a wide variety of spherically symmetric shock wave geometries in different dimensions. We find detailed agreement.

\medskip
\noindent
\end{abstract}
\end{titlepage}
\starttext \baselineskip=17.63pt \setcounter{footnote}{0}
\tableofcontents

\sc
\section{Introduction  }

The two sides of the Penrose diagram of an eternal AdS black hole are connected by an Einstein-Rosen  bridge (ERB). The ERB grows with time: classically it grows forever. On the other hand, the  dual boundary theories  very quickly come to thermal equilibrium. All evolution  seems to stop at the scrambling time $t_{\ast}.$ The scrambling time is a short time, only logarithmically greater than the light-transit-time across the black hole. This leads to a puzzle: if the quantum state of the two sides stops evolving, how can the continuing  growth of the ERB, over long periods of time, be described in the dual theory\footnote{Another way to put the question is: are there properties of the gauge theory wave function that can serve as clocks, and for how long can they continue to record the time? There are a number of more or less standard answers. The $\sim N^2$ part of the vertical entanglement (see section \ref{ERB length}) can be used as a clock, but it reaches its maximum after a multiple of the light-crossing time \cite{Hartman:2013qma}\cite{Liu:2013iza}. The decay of local correlation between the left and right  CFTs  may also be used \cite{Louko:2000tp}\cite{Kraus:2002iv}\cite{Fidkowski:2003nf}. These correlations exponentially decay for a time of order $S$, but then become noisy with an amplitude of order $e^{-S}.$ It is a special feature of quantum mechanics that there are properties of the wave function that are monotonic for much longer times.}?

The answer, of course, is that the quantum state does not stop evolving. Subtle quantum properties continue to equilibrate long after a system is scrambled. These properties can be summarized in a quantity called computational complexity, or just complexity. In the sense that we will use the term, complexity of  classical systems  cannot get very large.  Consider a system of $K$ classical bits  in the initial state $(00000....).$ Suppose our goal is to get to some other state. The  number of simple operations (one or two c-bit operations) required to accomplish the task will never be larger than $K.$  $K$ is also the maximum entropy of the c-bit system.

In quantum mechanics the situation is  different. Entropy is only the tip of a gigantic complexity iceberg. For $K$ qubits the maximum entropy is still $K.$ But for almost all states the number of 2-qubit gates needed to achieve the state is exponential in $K$ \cite{Knill}. Until recently this  difference between classical and quantum complexity has not played a large role in formulating physical principles, but this may be changing (see also \cite{Harlow:2013tf}).

In \cite{Susskind:2014rva}, a conjecture was made that the complexity of a state is proportional to the length of the ERB. Here we will refine this conjecture slightly: we propose that the total complexity, measured in gates, is
\be \label{conjecture1}
\c(t_L,t_R) = \frac{V(t_L,t_R)}{G_N l_{AdS}}
\ee
where $V$ is the spatial volume of the ERB. This volume is defined using the maximum volume codimension one surface bounded by the CFT spatial slices at times $t_L,t_R$ on the two boundaries. Equation (\ref{conjecture1}) should be understood up to an order one factor of proportionality, since we do not know how to define gate complexity more precisely than this.

A simple check on this proposal can be carried out using the time evolution of the thermofield double state $|TFD\rangle$ corresponding to the analytic eternal two-sided black hole \cite{Maldacena:2001kr}. In particular, in section \ref{ERBsection}, we will see that this formula has the right time dependence and scaling with temperature.

In section \ref{shocksection}, we will examine the conjecture is a less trivial setting. Building on \cite{Dray:1985yt}\cite{vanRaamsdonk}, references \cite{Shenker:2013pqa}\cite{Shenker:2013yza} constructed a wide class of shock wave geometries dual to perturbations of the thermofield double state. The complexity of such states is easy to estimate, so we are able to check (\ref{conjecture1}) for a large class of spherically symmetric states. A preliminary version of this check was carried out in \cite{Susskind:2014ira}.

The conjecture (\ref{conjecture1}) was partially inspired by the connection \cite{Hartman:2013qma} between time evolution, the length of the Einstein-Rosen bridges, and the tensor network description of quantum states. There should be a close connection between the circuit complexity defined in this paper and the minimal size of the tensor network description of a state. We will comment briefly about this after reviewing properties of complexity in section \ref{complexitysection}.

For the convenience of the reader we will list some assumptions conventions, notations that occur throughout the paper.

\bi
\item $D$ refers to the space-time dimension of the bulk theory.

\item We will often work in units such the AdS radius $l_{ads}$ is equal to unity.

\item Our discussion of precursors will be limited to the case of black holes with Schwarzschild radius of order the AdS scale $R \sim l_{AdS}$. For such black holes, the temperature of the black hole is of order $T\sim 1/l_{AdS}.$

\item Earlier papers \cite{Susskind:2014rva}\cite{Susskind:2014ira} focused on the length of the Einstein-Rosen bridge denoted by $d$. In this paper the focus is on the spatial volume of the ERB called $V$.

    For long, symmetric wormholes, $V$ and $d$ are related by the cross-sectional area $A$ of the ERB. The area is of order the entropy of the black hole in Planck units. Thus, up to factors of order $1$ we have
    $V =  A d = S  \ { l_p^{D-2}} \ d.$

\item Complexity can refer to an operator or to a state. In either case it is measured in gates. It is denoted by $\c.$

\item As usual, the two sides of the eternal black hole will be called left and right. This notation also refers to the two CFT's describing the boundaries.

\item The Killing time in a two-sided black hole is denoted $\tau.$ In the right side $\tau$ increases from past to future in the usual way. On the left side $\tau$ increases from future to past. In the Einstein-Rosen bridge $\tau$ is space-like and increases from left to right.

    \item The boundary time $t$ increases from past to future on both sides.

    \item The notation $r_m$ indicates a certain value of the (time-like) radial Schwarzschild coordinate inside the black hole where the function (to be defined ) $r^{D-2} \sqrt{|f(r)|}$ has a maximum.

		\item We will use $v_D$ to represent the value of this maximum: $v_D = \omega_{D-2}r_m^{D-2}\sqrt{|f(r_m)|}$ where $\omega_{D-2}$ is the volume of a unit $D-2$ sphere.

     \item   The symbol $t_{\ast}$ is used for the time $\frac{1}{2\pi T}\log S$. This is the scrambling time for black holes with $T\sim 1$.
		
		\item We use $t_f$ to denote the folded time interval associated to a state, defined below.
\ei

\section{Einstein-Rosen bridges}\label{ERBsection}
\subsection{Black hole geometry}


The metric of a Schwarzschild AdS black hole has the form

\be
ds^2 = -f(r) d\t^2 +f(r)^{-1} dr^2 + r^2 d\Omega_{D-2}^2
\label{metric}
\ee
where $f(r)$ is given by
\be
f(r) =r^2+1 -  \frac{\mu}{r^{D-3}}.
\label{fD}
\ee
For $D>3$, the parameter $\mu$ is determined by the mass as $\mu = 16\pi G_N M/(D-2)\omega_{D-2}$, with $\omega_{D-2}$ the volume of a $(D-2)$-sphere. In the $D = 3$ case corresponding to a BTZ black hole, the function is $f(r) = r^2 - 8G_N M$.

In equation (\ref{metric}) the time coordinate $\t$ runs from past to the future on the right side of the Penrose diagram and from future to past on the left side. We will introduce a boundary time $t$ which strictly runs from
past to future on both sides. Thus

\bea
t \eq \t  \ \ \ \ \ \ \ \ \rm right \ side \it  \cr \cr
t\eq-\t  \ \ \ \ \ \ \ \ \rm left \ side. \it
\eea

Let's review the boundary-bulk duality for wave functions. In the dual CFT   the system is described by a single state that depends on two times, $t_L, \ t_R.   $ The subscripts $L,  \ R$
represent left, right. The instantaneous state is written

$$|\Psi(t_L, t_R)\ra.$$
There are two commuting Hamiltonians that generate independent time-translations,

\bea
i\partial_{t_L} |\Psi\ra \eq H_L |\Psi\ra  \cr \cr
i\partial_{t_R} |\Psi\ra \eq H_R |\Psi\ra
\eea
The TFD is an eigenvector of $H_R - H_L$ with eigenvalue zero, but it evolves nontrivially with $H_R + H_L.$

\begin{figure}[h!]
\begin{center}
\includegraphics[scale=0.9]{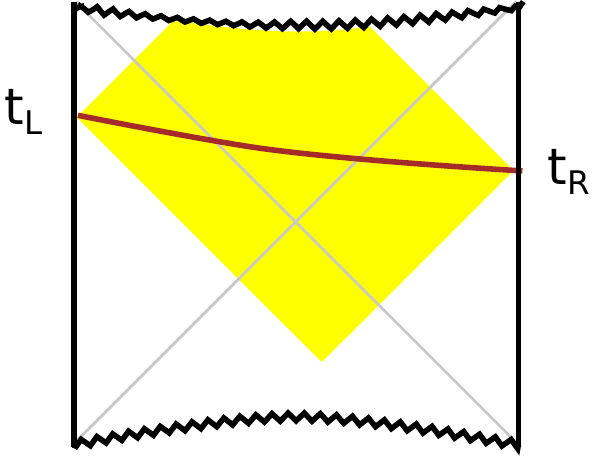}
\caption{The yellow region is the Wheeler-DeWitt patch for the times $t_L, t_R.$ The brown curve indicates a space-like
surface connecting the two boundaries.  }
\label{A}
\end{center}
\end{figure}

From the bulk viewpoint $|\Psi\ra$ represents a Wheeler DeWitt wave function covering a patch of the space-time geometry.  The patch contains
all spacelike surfaces which terminate on the  boundaries at times $t_L, \ t_R$ \cite{Maldacena:2013xja}. This is illustrated in figure \ref{A}.

We will be interested in the volume, $V(t_L, t_R),$ of the ERB that connects the two boundaries at the times $t_L$ and $t_R.$

\subsection{The size of an ERB }\label{ERB length}

The size of an  ERB is not a self-evident idea. In \cite{Susskind:2014rva} and \cite{Susskind:2014ira} a naive definition was proposed. The construction involved connecting the boundaries by $(D-2)$-dimensional surfaces similar to the Ryu-Takayanagi surfaces \cite{Ryu:2006bv}\cite{Hubeny:2007xt} used by Hartman and Maldacena \cite{Hartman:2013qma} to study the evolution of vertical entanglement\footnote{For the definition of vertical entanglement see \cite{Susskind:2014rva}\cite{Susskind:2014ira}.} In the special case of BTZ this reduces to geodesics connecting the two boundaries.

In  retrospect this was not a good idea for a number of reasons. One serious defect is that existence is not guaranteed. The extremal $(D-2)$ dimensional surfaces are saddle points, and it is possible to conceive of long asymmetric wormholes with no extremal surface connecting the two asymptotic regions. Concretely, because such surfaces are not topologically stable, they could slip off the transverse $(D-2)$-sphere when the ERB becomes long. Or they could run into the singularity. Also, the proposal fails to scale properly with temperature when the black hole is continued away from the Hawking-Page transition point.

There is, however, another simple candidate for the definition of the ERB size, which does not suffer from these problems. Consider the brown curve in figure \ref{A} connecting $t_L$ and $t_R.$ Taken literally, each point along the curve in the Penrose diagram represents a $(D-2)$-sphere. The curve itself is a $(D-1)$ dimensional spatial volume. Surfaces of this type fill the spatial volume of the ERB. Moreover, the extremal surfaces are maxima of the volume, not saddle points. For AdS black holes in Einstein gravity, we believe that the volume of such surfaces is always bounded from above (apart from a UV divergence near the boundary), so the existence of a maximum volume surface is guaranteed.\footnote{In Gauss-Bonnet, there is no global maximum, but for small Gauss-Bonnet parameter, there is still a local maximum.}

\subsubsection{Infinite time}
\begin{figure}[h!]
\begin{center}
\includegraphics[scale=.9]{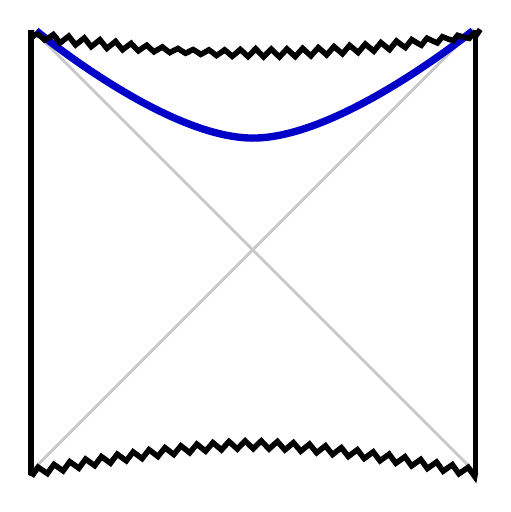}
\caption{Maximum volume surface for infinite $t_{L,R}.$ }
\label{erb}
\end{center}
\end{figure}

Consider the black hole geometry described in (\ref{metric}). The first step in constructing a maximal volume surface connecting $t_L$ and $t_R$ is to understand an especially simple limit in which the two boundary times are taken to infinity. This is illustrated in figure \ref{erb} where the blue curve represents the limiting configuration. The simplifying feature is the $\t$-translation symmetry of the geometry.
For the infinite-time ERB the volume extends over an infinite range of $\tau,$ and is therefore translationally invariant. Moreover the system is also rotationally invariant. It follows that the surface of maximum volume is located at a fixed value of $r.$ Its volume per unit $\tau$ is equal to

 \be
 \frac{dV}{d\t} = \omega_{D-2} r^{D-2}\sqrt{|f(r)|}.
 \ee

To find the maximal volume surface, we must maximize the RHS over $r$ between $r = 0$ and the horizon radius. From (\ref{fD}), it is easy to see that the function has a maximum at some $r_m$. We denote the maximum by $v_D$:
\be
v_D = \omega_{D-2} r_m^{D-2}\sqrt{|f(r_m)|}.\label{F}
\ee
For an AdS-scale black hole with $\mu \sim 1$, $r_m$ is also order unity. For a high temperature black hole $\mu \gg 1$, we find $r_m^{D-2}\sqrt{|f(r_m)|} = \mu/2$ and thus (restoring $l_{AdS}$)
\be\label{hight}
v_D = \frac{8\pi G_Nl_{AdS}}{D-2}M \hspace{20pt} (\text{high $T$}).
\ee

\subsubsection{Finite time}

In the appendix to this paper we present formulae for the volumes of symmetric ERBs connecting the boundaries at finite time, and we show how to compute them using a geodesic equation. Here we will give a simplified argument that captures the main features. The basic idea, articulated in a closely related setting by \cite{Hartman:2013qma}, is that if the ERB is long, the maximum volume surface tends to hug the surface defined by the infinite-time limit.  In the case of the unperturbed TFD-state it stays close to the infinite-time limit until $\t_L$ is approximately equal to $-t_L $ and $\t_R \approx t_R.$ It follows that the regularized volume of the bridge for $|t_L + t_R| \gg \beta$ is given by
\be
V(t_L, t_R) = v_{D} |t_L + t_R|   \ \  \ \ \ \ \ \rm ( large \ \it t_L+t_R \rm). \it
\label{grow linear}
\ee

Let us now perform a sanity check of the conjecture (\ref{conjecture1}). Using (\ref{hight}), and noting that $M \propto ST$, we find that the complexity of a high-temperature TFD state increases as
\be\label{tfdcomplex}
\c(t_L,t_R)   \propto S  T |t_L + t_R|
\ee
This is precisely the behavior one would expect based on a quantum circuit model of complexity \cite{Hayden:2007cs}\cite{Susskind:2014rva}: the rate of computation measured in gates per unit time is proportional to the product $ST.$ The entropy appears because it represents the width of the circuit and the temperature is an obvious choice for the local rate at which a particular qubit interacts.

We  note that $V$ is a function of $(t_L + t_R)$ as a consequence of the symmetry generated by the difference of Hamiltonians $(H_R-H_L).$ Time reversal symmetry of the TFD geometry implies that it should be an even function. For early time, i.e., $t_L+t_R \ll \beta,$ the volume is quadratic in $(t_L +t_R)$. A useful formula to have in mind is the length of geodesics in BTZ. These are not codimension-one surfaces, but the qualitative behavior of the length is similar to the higher-dimensional volume, and the exact formula can be worked out (here, for horizon radius $ = l_{AdS}$):
\be
d(t_L + t_R) = 2 \log{ \left[\cosh{\frac{1}{2}}(t_L +t_R) \right]}.
\label{d-cosh}
\ee

\section{Computational complexity}\label{complexitysection}
\subsection{Properties of complexity}\label{properties}

Although it is not essential, we will assume that the black hole can be modeled as a collection of $2K$ qubits. The Hilbert space of the two-sided system is ${\cal{H}}_L \times \cal{H}_R$ with each factor having dimension $2^K.$
 We take $K=S.$

Any unitary operator $V$ in the $2^K$ dimensional Hilbert space of the left-side black hole has a computational complexity $\c_V,$ which was defined\footnote{Patrick Hayden has pointed out that a better definition of complexity may be the minimal depth of a quantum circuit, rather than the minimum number of gates, needed to generate $V.$ The minimum depth would  be identified with the complexity per qubit. There are some cases where it is important to use circuit-depth rather than number of gates. An example is the complexity needed to scramble. There are circuits with only $K$ gates which can scramble. However, despite the small number of gates the depth of these circuits is $\log K.$ This subtlety is not important in this paper.

A smoother but closely related notion of complexity was defined by Nielsen and collaborators \cite{Dowling}, based on geodesic length in a Riemannian ``complexity geometry." We thank Nathaniel Thomas for helpful explanations of complexity geometry.}
to be the number of  2-qubit gates in the smallest quantum circuit that approximately generates $V$ \cite{Susskind:2013aaa}\cite{Susskind:2014rva}\cite{Susskind:2014ira}. For example the unit operator has zero complexity. Simple products of all the qubits have complexity of order $K.$
Operators of complexity $\c = K \log K$ can scramble an initial product state.

Quantum complexity  can grow far beyond the scrambling complexity. However it is bounded by an exponential of $K.$
The maximum complexity satisfies

\be
\log \c_{max} \sim K.
\label{cmax}
\ee

Generating an operator with exponential complexity requires an exponential time. But in a technical sense, exponential complexity is not rare. It can be shown \cite{Knill} that almost all unitary operators have complexity satisfying (\ref{cmax}). Nevertheless, our discussion will be restricted to shorter time-scales during which the complexity is much smaller than maximal.

If $U_L(t)$ and $U_R(t)$ are the time evolution operators for the two sides, their complexity will increase with $|t|.$ Typically, apart from a short transient at early time, the complexity will grow linearly, the proportionality factor being $K$ or the entropy.

\be
\c(t) = Kt.
\label{linear}
\ee
We expect this behavior to continue until the complexity reaches $\c_{\max}$.
Then it fluctuates around $\c_{max}$ for an extremely long, doubly exponential, quantum recurrence time.

So far, we have discussed the complexity of operators. We can also define the complexity of a quantum state. To do that we need to define a fiducial state of zero complexity. For the $2K$ qubit system, we can take it to be the state

$$|0\ra \equiv |00000000...00\ra$$
The complexity of a general state $|\psi\ra$ is defined as the complexity of the least complex unitary operator which will give $|\psi\ra$ when applied to $|0\ra.$

\begin{figure}[h!]
\begin{center}
\includegraphics[scale=.7]{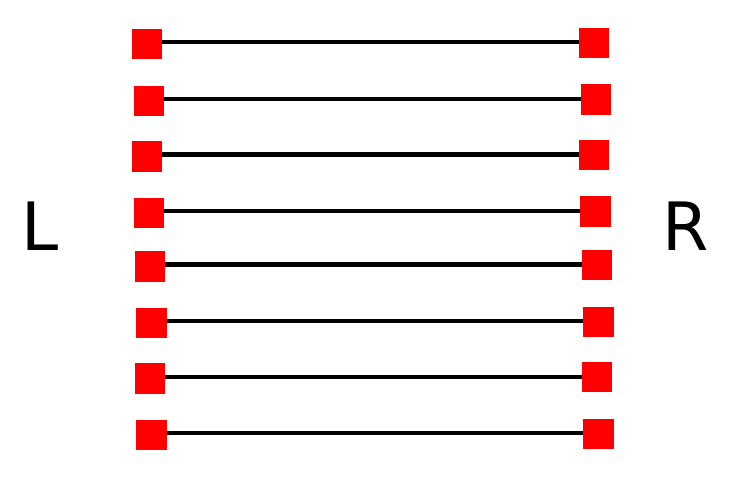}
\caption{qubit model for the TFD state. The TFD consists of a product of Bell pairs shared between the
left and right sides. }
\label{TFD}
\end{center}
\end{figure}

The TFD state is close to being maximally entangled. To an approximation that we will discuss later, it can be identified with a product of Bell pairs in which each pair is shared between the left and right systems. This is illustrated in figure \ref{TFD}.
 It takes one gate to act on a pair of qubits in the state $|00\ra$ to turn it into a Bell pair. To create $K$ Bell pairs requires $K$ gates. Therefore the complexity of the TFD is $K$.

After an early-time transient period, the 
 complexity of the state

\be
|\Psi(t_L, t_R) \ra = U(t_L) U(t_R)|TFD\ra
\ee
will grow like

\be
\c(t_L,t_R) = K|t_L + t_R|
\label{c-form}
\ee
until it becomes maximal. Identifying the temperature as the conversion between time in the CFT and time in the analog qubit system, we find that this agrees with the complexity derived from the volume of the ERB in (\ref{tfdcomplex}). The numerical coefficient of proportionality is ambiguous (and possiby dimension-dependent), because complexity itself is ambiguous up to a numerical factor. However, the relative normalization of complexity and ERB volume can be  fixed once and for all  by comparing (\ref{c-form}) and (\ref{grow linear}). We will use this below.

\subsection{Complexity of a precursor}\label{precursorSec}

In this section, and in most of the rest of the paper, we will restrict our attention to black holes with temperature of order the AdS scale.

\begin{figure}[h!]
\begin{center}
\includegraphics[scale=.3]{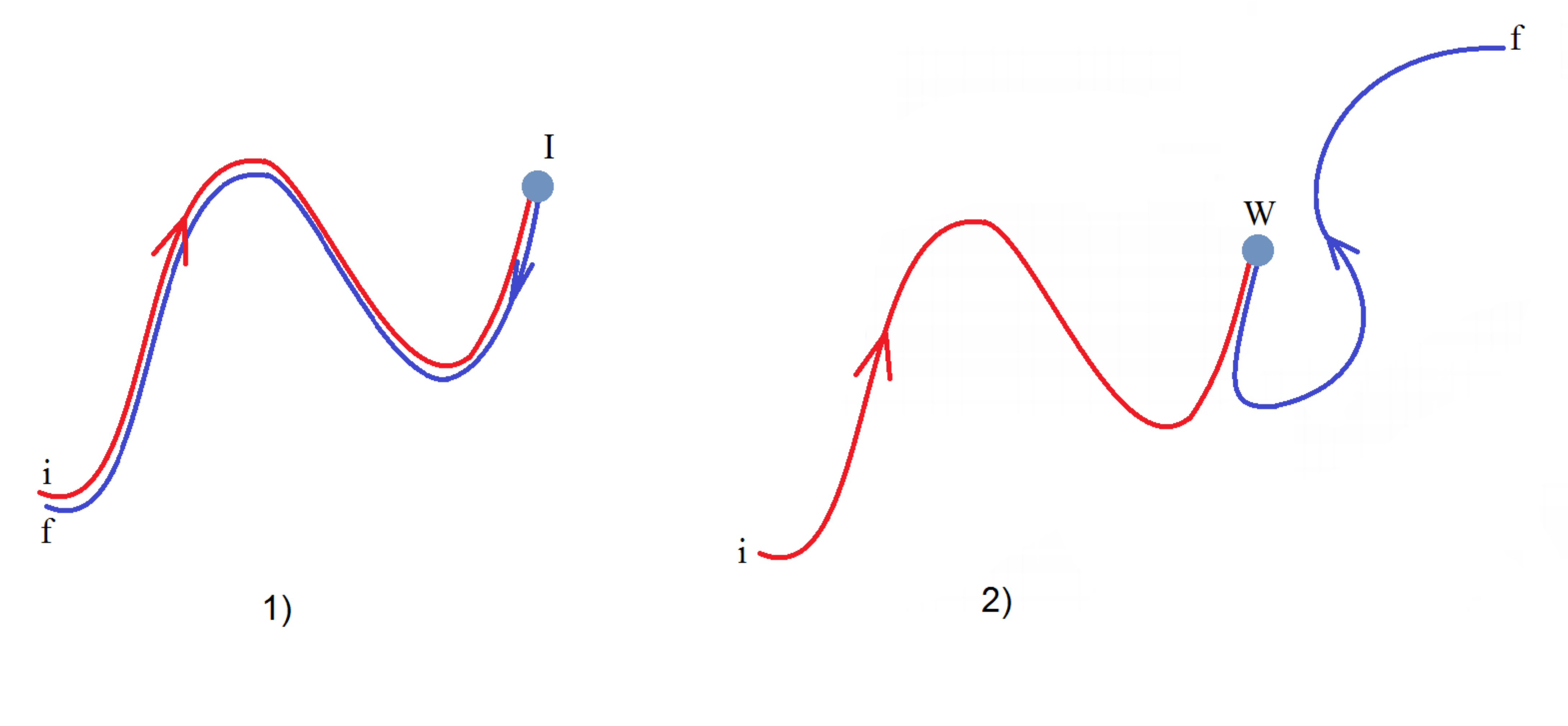}
\caption{In the left panel the operation  $U^\dag(t) I U$ is illustrated. The letters $i$ and $f$ represent initial and final states. The backtracking trajectories illustrate the cancelation of $U$ and $U^{\dag}.$  In the right panel the unit insertion is replaced by the insertion of $W.$ The backtracking of trajectories takes place for a limited time until the butterfly effect kicks in at the scrambling time. }
\label{butterfly}
\end{center}
\end{figure}

A precursor is a unitary operator of the form,

\be
W(t) = U^{\dag}(t)WU(t).
\label{precursor}
\ee
Although it has the form of a Heisenberg operator, we think of it as an operator in the Schrodinger picture. As $t$ grows, either positively or negatively, the complexity of $W(t)$ grows. This requires some explanation. If $W$ is the unit operator then the complexity does not grow since the $U$ and $U^{\dag} $
cancel. In \cite{Susskind:2014rva} it was explained that the chaotic nature of the dynamics destroys the cancelation when an operator $W$ is inserted even if $W$ itself is very simple; say a single qubit. The reason of course is the butterfly effect. This was illustrated in figure 1 in  \cite{Susskind:2014rva} which we reproduce as figure \ref{butterfly}. The insertion of $W$ disrupts the time-reversed evolution described by $U^{\dag}$ and quickly causes the trajectories to diverge. For this reason it was argued that for $t\gg t_{\ast}$ the complexity of $W(t)$ is just twice the complexity of $U(t).$ A more refined guess would take into account the partial cancelation that should take place until the butterfly effect kicks in. As we will see the time-scale for the delay is the scrambling time. Therefore a refinement of the estimate would be that the 
complexity of $W(t)$ for $t > t_*$ is given (to accuracy of order $K$) by

\be
\c_W(t) = 2K(t - t_*).
\label{cancel}
\ee

We can illustrate the behavior in (\ref{cancel}) in the Hayden-Preskill circuit model \cite{Hayden:2007cs}. The reader is referred to \cite{Susskind:2014ira} for the details of the model. Let's begin with a very simple state of $K$ qubits, namely

\be
|\psi(0)\ra = |00000....\ra.
\ee
We focus on a particular qubit labeled $W_a.$ After $n< \log_2 K$ parallel time steps the  qubit $W_a$ will have interacted, either directly or indirectly, with $2^n$ qubits. Let us call that subset $A.$ The remaining $K-2^n$ qubits have had no contact with $W_a.$  The evolution operator is a product of gates. It factors into an operator for the subset $A$ and another factor for the complement of $A,$ which we call $B.$

\be
|\psi(n)\ra = U_B(n) U_A(n)|00000....\ra.
\ee
The  operators $U_A$ and $U_B$ are built out of non-overlapping sets of qubits and commute with each other.

Next act with the qubit operator $W_a.$ Then run the system back with the operator $U_A^{\dag}U_B^{\dag}.$ The $B$ operators cancel and the result is

\be
|\psi(2n)\ra = U_A^{\dag}(n)W_a U_A(n)|00000....\ra.
\ee
This resulting state is a tensor product of $A$ and $B$ states. The $A$ factor is scrambled, but the $B$ factor has all qubits
in the state $0.$ For example when $n = \log_2 K-3$ at least seven-eights of the qubits are unaffected by the evolution. Obviously not much complexity has been generated during this time. However as soon as $n=\log_2 K$ the entire system  becomes fairly scrambled. This happens rather suddenly. Once that point has passed, the complexity begins to grow linearly with time. Thus we see that the growth of complexity is delayed   by  $2\log_2 K,$ i.e., twice the scrambling time.  This is the circuit analog of (\ref{cancel}).

\subsection{A note on tensor networks}

Hartman and Maldacena \cite{Hartman:2013qma} have proposed a tensor network (TN) picture to illustrate the evolution of ERBs.\footnote{The Hartman-Maldacena tensor networks cannot resolve distances on scales smaller than $l_{ads}.$ The tensors do not act in a space of a single qubit but rather on the entire Hilbert space of an $N \times N$ matrix theory. The TN picture makes the most sense for black holes of radius $R\gg l_{ads}.$}

\begin{figure}[h!]
\begin{center}
\includegraphics[scale=.33]{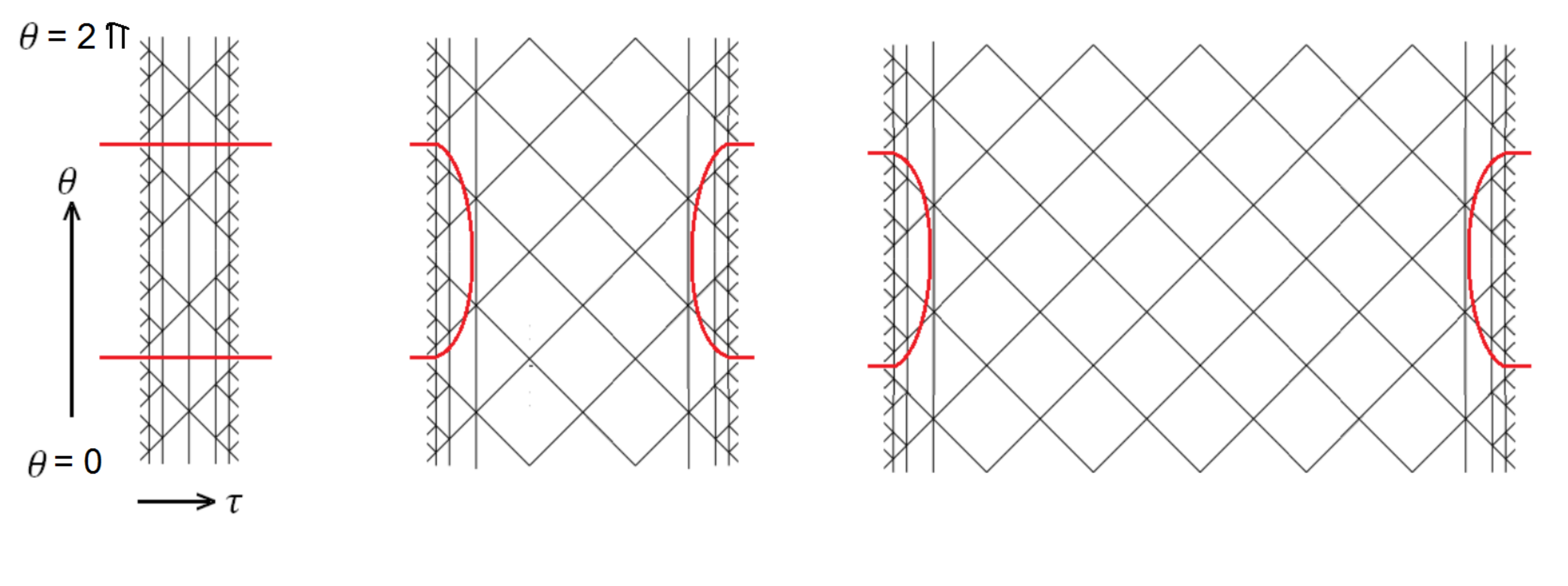}
\caption{Evolution of the ERB tensor network. The red curves depict the RT surface for computing vertical entanglement. The tensor network fills the volume of the ERB.}
\label{TN}
\end{center}
\end{figure}

Figure \ref{TN}
shows the evolution of the Hartman-Maldacena TN for the  ERB in the case of a 1+1 dimensional boundary theory (D = 3). As time increases, more layers along the $\tau $ direction are added to the TN.  Near the left and right boundaries the network shows the familiar scale-invariant pattern of associated with geometry of AdS \cite{Swingle:2009bg}. At the horizon the pattern changes to reflect the $\t$-translation invariance of a long ERB. The wave function of the boundary state, obtained by contracting all the internal indices in the TN, evolves because the network keeps growing. In many ways the evolution of the network resembles the evolution of a quantum circuit, the width of the circuit being the number of layers in the $\theta $ direction and the depth being the number of layers in the $\tau$ direction.

  For a given boundary state the associated TN is not unique (this is also true of quantum circuits).
  It is tempting to identify the complexity of  the boundary state  with the number of nodes of the smallest tensor network that can generate the state.

There is an upper limit on the complexity of a state of a system of $K$ qubits, exponential in $K.$  What happens when the depth of the tensor network exceeds some exponential length? The bounded nature of complexity implies that the state generated by the TN can be constructed by a smaller TN. In some sense there is an upper bound on the size of the TN. This implies a breakdown in the classical geometric description of an ERB when the time becomes exponential in the entropy $S.$

\section{Complexity and shock wave geometries}\label{shocksection}

In what follows we will test the conjecture $\c \propto V$ by considering evolutions more general  than those generated by $H_L$ and $H_R.$ In particular we consider perturbed geometries generated by applying thermal-scale operators $W_L(t_L)$ as discussed in \cite{Shenker:2013pqa}\cite{Shenker:2013yza}. In the qubit model the $W$-operators may be thought of as one-qubit traceless Pauli operators with complexity of order unity.

\begin{figure}[h!]
\begin{center}
\includegraphics[scale=.9]{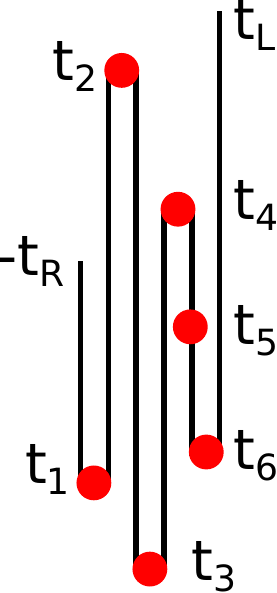}
\caption{Time-fold with six insertions. The insertions at $t_1, t_2, t_3, t_4 \rm and \it t_6$ occur at switchback points.
The insertion at $t_5 $ does not. }
\label{B}
\end{center}
\end{figure}

Let $t_1, t_2, t_3,..., t_n$ be a series of left-side times, typically not in time-order. We consider the state
\be
|\Psi(t_L, t_R) \ra = U_R(t_R) U_L(t_L) W_L(t_n)W_L(t_{n-1}).....W_L(t_1)|TFD\ra.
\ee
Using the fact that $H_L - H_R$ annihilates $|TFD\rangle$, we can re-write this using $L$ operators only, as
\be\label{perturbedstate}
|\Psi(t_L, t_R) \ra = U_L(t_L) W_L(t_n)W_L(t_{n-1}).....W_L(t_1)U_L^\dag(-t_R)|TFD\ra.
\ee
Since the operators are generally out of time order, the evolution can be represented by a time-fold \cite{Heemskerk:2012mn}\cite{Susskind:2013lpa} as shown in figure \ref{B}.
Note that there are two kinds of insertions illustrated in figure \ref{B}. Some insertions---the ones at $t_1, t_2, t_3, t_4 $ and $t_6$---occur at fold points or ``switchbacks." Others like the one at $t_5$ are ``through-going."

This time fold diagram represents a recipe for making the state $|\Psi(t_L,t_R)\rangle$: beginning with the TFD, evolve forwards and backwards with the Hamiltonian, inserting local operators at the locations of the red dots. This recipe gives us an upper bound on the complexity of the state $|\Psi(t_L,t_R)\rangle$, namely the total folded time interval $t_f$:

\be
\frac{\c(\Psi)}{K}\le t_f \equiv |t_1+t_R| + |t_2-t_1| + |t_3-t_2|+......+ |t_L - t_n|.
\ee

How tight do we expect this bound to be? If the time $t_f$ is less than exponential in the entropy, we expect the bound to be tight, except for the partial cancelations described in section \ref{precursorSec}. These cancelations occur at each switchback. We therefore have

\be\label{conjform}
\frac{\c(\Psi)}{K} = t_f - 2 n_{sb}t_*
\ee
where $n_{sb}$ is the number of switchbacks, and we have assumed that $|t_i - t_{i-1}| > t_*$.

In the next section, we will use this formula to check the conjecture of \cite{Susskind:2013aaa}\cite{Susskind:2014rva}\cite{Susskind:2014ira}. This conjecture, with our refinement, states that the complexity is proportional to the volume of the ERB. References \cite{Shenker:2013pqa}\cite{Shenker:2013yza} showed how to construct geometries dual to perturbed TFD states (\ref{perturbedstate}), so checking $\c \propto V$ amounts to a concrete maximal surface problem in these geometries. Before we begin this calculation, we will emphasize three points:
\begin{itemize}
\item  We can normalize the relationship between complexity and volume using the pure TFD state as discussed in section ~\ref{properties}. This allows us to check the agreement for states of the form (\ref{perturbedstate}) including the coefficient of proportionality.
\item We are restricting our attention to black holes with temperature of order the AdS scale. For such black holes, thermal scale perturbations are approximately as large as the entire system. This partially justifies our use of spherically symmetric shock wave geometries. It is simple to generalize (\ref{conjform}) for localized perturbations of a large spatially extended system. However, the corresponding maximal surface problem becomes significantly harder, and will be left to future work \cite{tobecontinued}.
\item We will verify the relationship $\c \propto V$ for large $|t_i - t_{i-1}|$, ignoring corrections that are $O(1)$ in the time differences, i.e. $O(K)$ in terms of the total complexity. We will, however, retain terms involving $t_*$, which are $\sim K \log K$.
\end{itemize}

\sc
\subsection{Finding the maximal surface}
\begin{figure}[ht]
\begin{center}
\includegraphics[scale = .7]{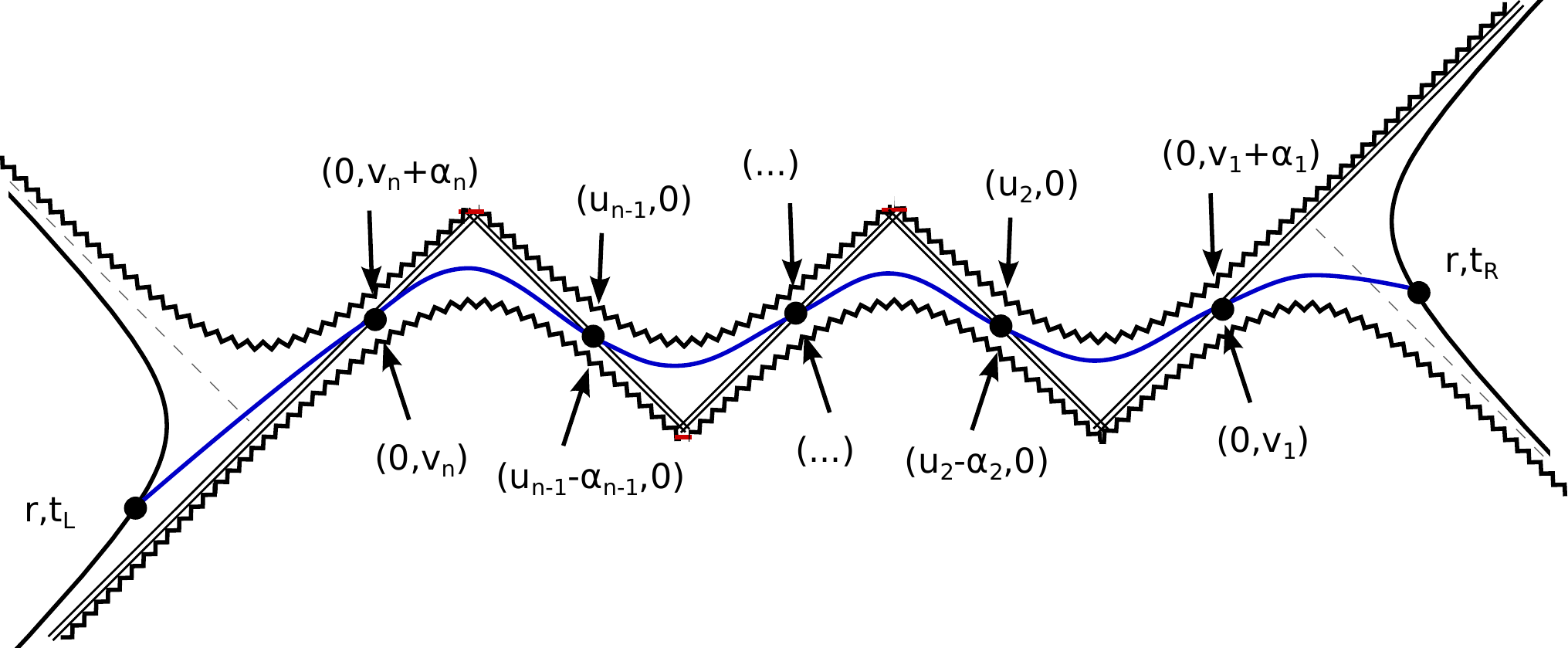}
\caption{Kruskal diagram for an ERB dual to the TFD state perturbed by out-of-time-order operators at the left boundary. The blue curve represents the maximal-volume surface crossing the ERB. Each point of intersection with a shock has two sets of $(u,v)$ coordinates: one in the patch the left, and one in the patch to the right. These are related by null shifts determined by the strength of each shock.}\label{folded}
\end{center}
\end{figure}

It is convenient to write the metric (\ref{metric}) in Kruskal coordinates,
\begin{align}\label{eternal}
ds^2&=-\frac{4 f(r)}{f'(R)^2}e^{-f'(R)r_*(r)}dudv + r^2 d\Omega_{D-2}^2\\
uv &= -e^{f'(R) r_*(r)} \hspace{20pt} u/v = -e^{-f'(R)t},\label{defofkruskal}
\end{align}
where $R$ is the horizon radius and $dr_* = f^{-1}dr$ is a tortoise coordinate. Adding a thermal-scale perturbation far in the past on the left leads to a null shock wave along the $u = 0$ horizon \cite{Shenker:2013pqa}. Adding a perturbation far in the future leads to a shock along $v = 0$. A sequence of perturbations as in (\ref{perturbedstate}) leads to a geometry with a long wormhole crossed by intersecting null shocks. These geometries were worked out in \cite{Shenker:2013yza}. A folded time axis with $n$ folds leads to a geometry with $n$ alternating shocks. A sample Kruskal diagram is shown in figure \ref{folded}.\footnote{Notice that we are sending in all shocks along either $u = 0$ or $v = 0$. In reality, they will be at finite $u,v$, but if the relative boost between adjacent shocks is large, we can take them to be along the horizon.}

This geometry is obtained by pasting together portions of the eternal AdS black hole metric across the horizons $u = 0$ or $v = 0$ with null shifts in the $v$ or $u$ directions of magnitude
\be
\alpha_i=2\exp{\left[ -\frac{2\pi}{\beta}(t_*\pm t_i)\right]}
\ee
Here, the sign depends on whether the shock is left-moving or right-moving. We will take this as the precise definition of $t_*$, but we note that it leads to $t_* \approx \frac{\beta}{2\pi}\log N^2$.

A maximal volume surface connecting $t_L$ and $t_R$ is also drawn in figure \ref{folded}. It is formed from $n+1$ pieces of maximal surface in the unperturbed geometry, connected at the $n$ locations where the surface crosses a shock. To each intersection point, we assign two different Kruskal coordinates. One is the location in the Kruskal system to the right of the shock, and the other is the location in the Kruskal system to the left. These are related by null shifts of magnitude $\alpha_i$.

Let us begin by considering a folded time axis with only switchback insertions, i.e. no through-going insertions. To keep the notation simple, we will also focus on a specific case with an odd number $n$ of total insertions, with $t_1 < -t_R$ and $t_n < t_L$. This is the case drawn in figure \ref{folded}. For odd $i$ we have the ``+'' sign in the definition of $\alpha$, and for even $i$ we have the ``-'' sign. The combined volume of the $n+1$ segments is
\begin{align}
V = V(t_R,v_1) + V(v_1+\alpha,u_2) + ... + V(u_{n-1} - \alpha_{n-1},v_n) + V(t_L,v_n+\alpha_n).
\end{align}
Using the formulas of appendix \ref{appendix}, and assuming all volumes are large (i.e. $|t_{i+1}-t_i|>t_*$), we find
\begin{align}
\frac{2\pi V}{\beta v_D} = &\log(- v_1e^{-2\pi t_R/\beta}) + \log\big[u_2(v_1+\alpha_1)\big]+... \\ &+\log\big[v_n(u_{n-1} - \alpha_{n-1})\big]+\log\big[(v_n+\alpha_n)e^{ 2\pi t_L/\beta}\big] + O(1).
\end{align}
Here and below, $O(1)$ represents a contribution that is $O(1)$ in terms of the various time variables. To ensure that the piecewise-maximal surface is actually maximal, we extremize this formula over the intersection points. This leads to $v_i = -\alpha_i/2$ for odd $i$ and $u_i = \alpha_i/2$ for even $i$. Plugging in the definition of $\alpha_i$, we find
\be\label{int}
V = v_D(-t_R - 2t_1 + 2t_2 - ... + -2t_n + t_L - 2nt_*) + O(1).
\ee
For the configuration of times specified above, this is simply
\be \label{volumeresult}
V = v_D (t_f - 2 n t_*) + O(1)
\ee
in precise agreement with the conjecture (\ref{conjecture1}) and the formula (\ref{conjform}). Different configurations of the times $\{t_i\}$ lead to different $\pm$ assignments for $\alpha_i$. The formula (\ref{int}) changes, but (\ref{volumeresult}) remains valid.

What is the effect of through-going insertions? We have seen that the shocks associated to switchback insertions already lead to an ERB volume that agrees with the complexity. In order for $\c \propto V$ to be correct, through-going insertions must not significantly change the volume of the ERB.

In fact, this is the case. The shock sourced by a through-going insertion insertion will run parallel to the shock associated with one of the adjacent switchback points; its only effect will be to slightly increase the strength of this adjacent shock. Let us illustrate this in the simplest case, with two shocks at times $t_1,t_2$. Suppose that we want to compute the volume of the bridge at $t_L = t_R = 0$, and the times satisfy $t_1 < t_2 < 0$.
\begin{figure}[ht]
\begin{center}
\includegraphics[scale = .65]{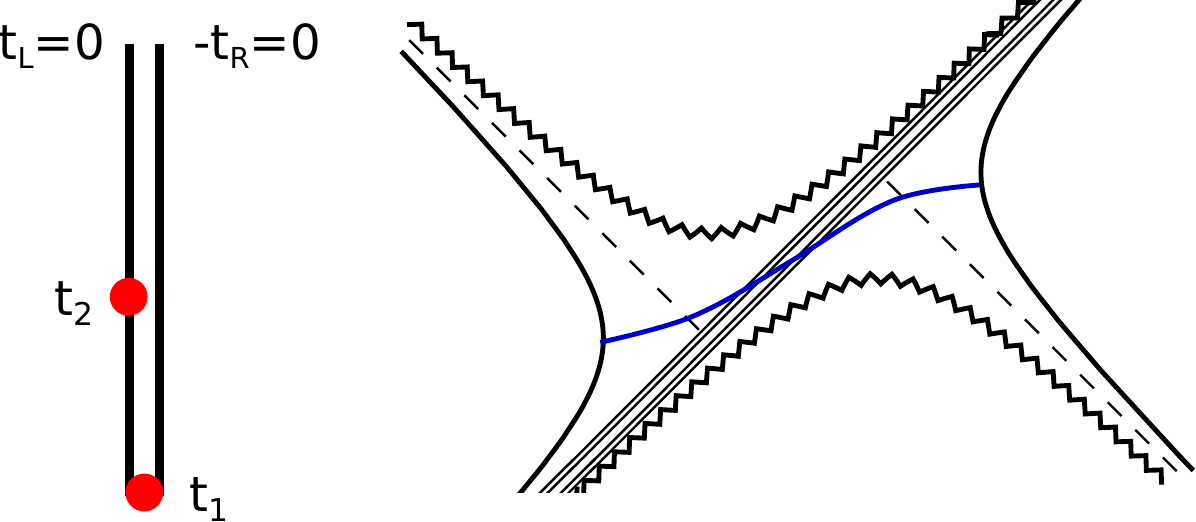}
\caption{Folded time axis and geometry for two shocks with $t_1< t_2 < t_L$. The $t_2$ insertion is through-going, and the associated shock runs right beside the stronger $t_1$ shock.}\label{twotwo}
\end{center}
\end{figure}
In the region that the maximal volume surface crosses the shocks, and in the boost frame appropriate to that surface, the shocks are very close together and running parallel. Parallel shocks superpose, so we can simply add the metrics together, obtaining a single effective shock with null shift $\alpha_1 + \alpha_2$. The volume is therefore proportional to $\log (\alpha_1 + \alpha_2) = \log \alpha_1 + O(1)$. The effect of the through-going insertion is to therefore to increase the complexity by a small amount, at most of order $K$.

\subsection{A maximal entanglement reflection principle}\label{reflection}

There are  features of the behavior of $V(t_L,t_R)$ which surprised us when we first discovered them.  They also happen to be features of the geodesic distance $d(t_L, t_R) $ for BTZ black holes. In the case of geodesic distance we can write down a simple analytic formulas for $d(t_L, t_R)$ in shock wave geometries. We will illustrate the points for the case of a single shock wave created at (negative) time $t_1.$
 \be
d(t_L, t_R) = 1 + 2 \log \left[ \cosh \frac{t_L+ t_R}{2}   + qe^{(2|t_1| +t_L -t_R)/2} \right].
\label{d(tL,tR)}
\ee

The first point is seen by setting $t_L=t_R.$ We note that the result is an even function of $t_L+t_R,$ i.e.,
\be
d(t_L+t_R) = d(-t_L -t_R).
\ee
The second surprising feature can be seen by fixing $t_L,$ say at $t_L=0,$ and noting that $d(0,t_R)$ decreases with $t_R$ for a fairly long period of time. The same two features are also found in the volume function $V(t_L,t_R).$

This first feature is  surprising because the insertion of the shock wave at $t_1$ explicitly breaks the time-reversal symmetry. It is not obvious why the complexity should be an even function of $(t_L + t_R).$ The second feature is even more surprising when we interpret $V$ as complexity; why should the complexity decrease as a function of $t_R?$

\begin{figure}[h!]
\begin{center}
\includegraphics[scale=.7]{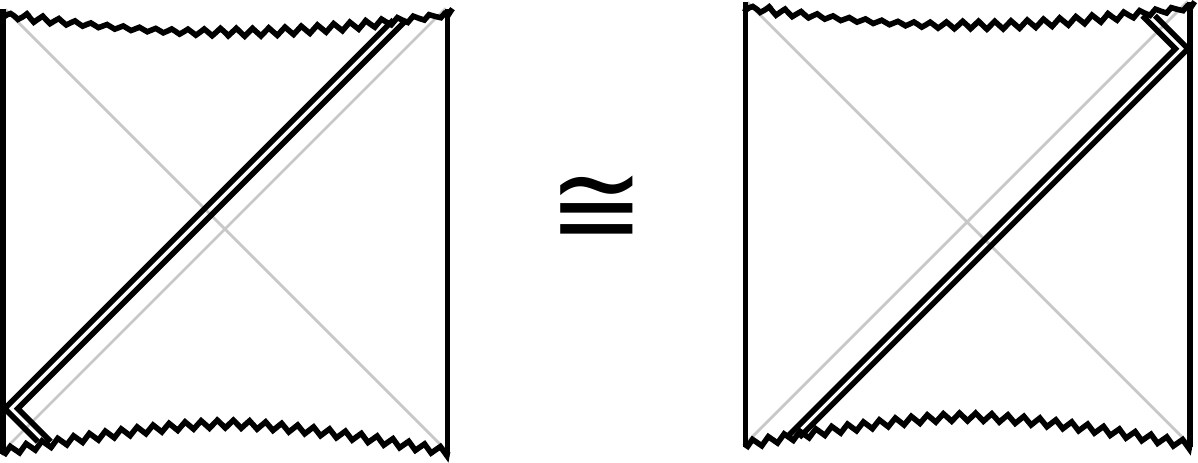}
\caption{For a maximally entangled state there is an equivalence between acting with unitary operators
on the left and right. The two shock wave geometries shown in the figure would be equivalent.}
\label{D}
\end{center}
\end{figure}

Neither of these features are accidental. They are  related to properties  of the TFD state. The maximally entangled model for the TFD is a product of $K$ Bell pairs. Such a state has the property that acting with any unitary operator on the left side is equivalent to acting with a reflected operator on the right side. Thus, if the TFD were maximally entangled, acting  with $W_L$ on the left at $t=0$,  would be equivalent to acting with the corresponding $W_R$ on the right side at $t=0:$
\be
W_L(t=0) |TFD\ra = W_R(t=0) |TFD\ra.
\ee
If we use the symmetry of $|TFD\ra$ under transformations generated by $H_R - H_L$ we can generalize this to
\be
W_L(t_1) |TFD\ra = W_R(-t_1) |TFD\ra.
\label{reflect}
\ee
This ``reflection principle" is illustrated in figure \ref{D}. 

The TFD state is not exactly maximally entangled, but for shock wave geometries generated by very low-mass perturbations with very large time separations between, it seems that maximal entanglement is a good approximation. From the bulk geometry, this is clear: in figure \ref{D}, if we focus on the geometry near the $t = 0$ slice, very early shocks sent in from the left are almost indistinguishable from very late shocks sent in from the right.

Note that the formula (\ref{reflect}) can be used to move operators from $L$ to $R$, or vice versa, in multi-shock states. This means that all multi-shock states can be represented, in the approximation described above, in terms of perturbations purely on the left. This is why we have focused on such perturbations throughout the paper.

Let us consider the two surprising features described above, in light of the reflection principle. First, take the state
\be
U_R(t)U_L(t) W_L(t_1) |TFD\ra
\label{state t}
\ee
corresponding to $t_L = t_R = t.$ Using \ref{reflect}, we write
\be
U_R(t)U_L(t) W_L(t_1) |TFD\ra = U_R(t)U_L(t) W_R(-t_1) |TFD\ra.
\label{symmetry}
\ee
By time reversal and left-right interchange this is equal to
\be
U_R(-t)U_L(-t) W_L(t_1) |TFD\ra.
\label{state -t}
\ee
Thus comparing \ref{state t} and \ref{state -t} it would follow that $V$ is a symmetric function of $t$ even though the time reversal symmetry is broken by the insertion of $W(t_1).$

Now let us consider the second feature: the decrease of complexity with increasing  $t_R$ for a  certain period of time. According to (\ref{reflect}) the state $U_R(t_R) W_L(t_1)|TFD\ra$ satisfies
\be
U_R(t_R) W_L(t_1)|TFD\ra = U_R(t_R) W_R(-t_1)|TFD\ra.
\ee
A left-right flip and a time reversal relates this state to
\be
 U_R(t_R) W_R(-t_1)|TFD\ra  \to U_L(-t_R) W_L(t_1)|TFD\ra.
\ee
The decrease of complexity with $t_R$ in the  state  $U(t_R) W_L(t_1)|TFD\ra$ is thus mapped to an increase of complexity with $t_L$ in the flipped state. This increase with $t_L,$ following the action of $W_L(t_1),$ is the expected behavior.\footnote{Note that the decrease with $t_R$ only lasts until  $t_R = |t_1|-2t_*.$ One can see an example of this in \ref{d(tL,tR)}. At $t_R = |t_1|-2t_*$ there is a crossover between the two terms and $d$ begins to increase with $t_R.$ This is also the expected behavior from the reflection principle.}

\sc
\section{Conclusion}
Quantum computational complexity---thought of as a property of the state of a system---is an extremely subtle quantity; given a state, there are very few tools to compute its complexity. All  the ordinary quantities that we are familiar with stop evolving by the scrambling time when the system reaches local equilibrium.  Nevertheless the computational complexity of a state is well defined and continues to  increase long after ordinary equilibrium is reached. It only  saturates at the classical recurrence time $ \sim e^S. $

For ordinary purposes computational complexity is far too subtle to be relevant for any real experiment on a chaotic system.  However it appears to play a fundamental role in encoding properties of the interiors of black holes. More generally it may be important for describing phenomena behind any event horizon, including cosmic horizons.

The assumption that ERB volume, $V(t_L, t_R),$ is determined by the complexity of the dual CFT state, together with some assumptions about the growth of complexity with time leads to a detailed conjecture for how $V(t_L, t_R)$
behaves in spherically symmetric shock wave geometries. This conjecture was checked for all such geometries in all dimensions.

One might wonder whether the equivalence between folded time and ERB volume is simply a geometric fact having nothing to do with chaos and complexity. The smoking gun implicating these properties is the partial cancelation occurring at switchback points. Quantum circuits allow us to see that the complexity of a precursor $W(t)$ is overestimated by the sum of the complexities of the evolution operators in \ref{precursor}. The Hayden Preskill circuit model \cite{Hayden:2007cs} gives a precise value for the overestimate \ref{cancel}. The value  agrees with our guess in \ref{cancel}, and more importantly, it agrees with the calculation of ERB volumes in shock wave geometries.

The occurrence of the scrambling time in the formula is a clear indication that the effect is connected with chaos and complexity. The fact that the same cancelation occurs---in just the right way---for the length of ERBs is quite remarkable.

\section*{Acknowledgements}

We are grateful to Patrick Hayden and Steve Shenker for discussions. We thank Dan Roberts for drawing our attention to a difficulty in reconciling the area of codimension two surfaces and the complexity of localized perturbations of the TFD state \cite{tobecontinued}.  This was part of our motivation for considering codimension one surfaces.

Support for this research came through NSF grant Phy-1316699 and the Stanford Institute for Theoretical Physics.

\appendix

\section{Maximal volume surfaces in the eternal black hole}\label{appendix}
\begin{figure}[ht]
\begin{center}
\includegraphics[scale = .7]{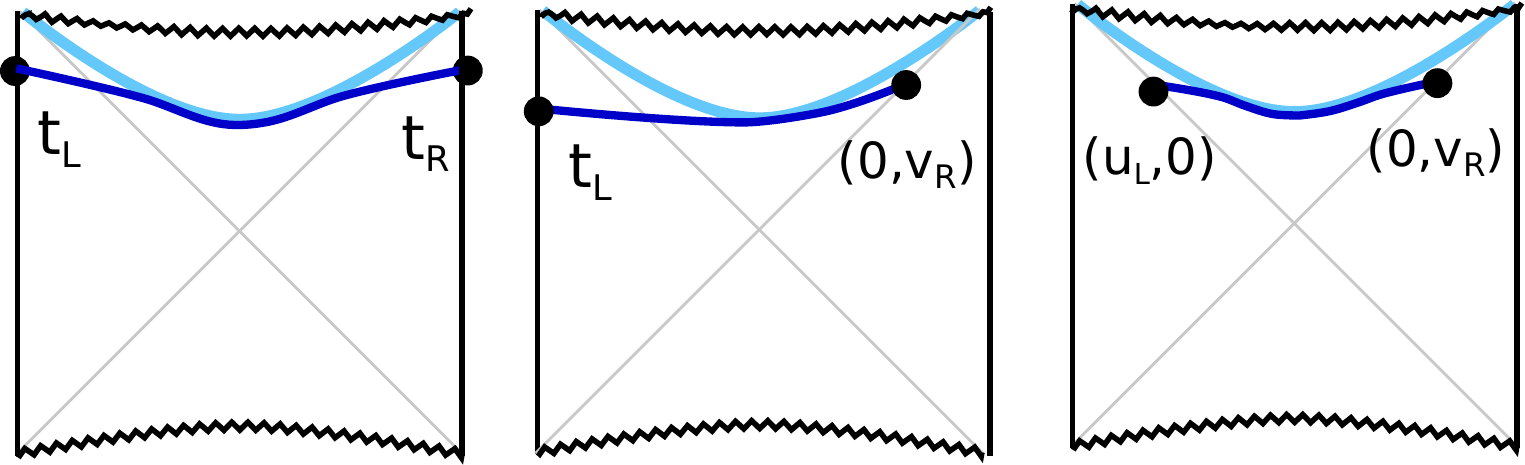}
\caption{The dark blue curves are the maximal surfaces defining $V(t_L,t_R)$, $V(t_L,v_R)$, and $V(u_L,v_R)$. The pale blue curve is the limiting infinite-time maximal surface. }\label{higherd}
\end{center}
\end{figure}

In this appendix, we will present formulas for the volumes of maximal surfaces in the unperturbed AdS black hole. The analysis in the main text required the lengths of three types of surfaces, shown in figure \ref{higherd}.

The volumes of the surfaces shown in the left and center panel are divergent, because of the infinite volume near the boundary of AdS. As usual, we define regularized volumes by subtracting an infinite but state-independent constant. The results, when the regularized volumes are large, are
\begin{align}
V(t_L,t_R) &= v_D |t_L + t_R| + O(1) \label{HMworkedout}\\
V(t_l, v_R) &= \frac{\beta}{2\pi}v_D\log(v_Re^{2\pi t_L/\beta})+ O(1) \\
V(u_L,v_R) &=  \frac{\beta}{2\pi}v_D\log(u_Lv_R) + O(1),\label{tocheck}
\end{align}
where $O(1)$ represents a contribution which does not increase as the volume becomes large. The quantity $v_D$ was defined in (\ref{F}). We will illustrate in detail the derivation of the third equation. Very similar computations (for codimension two surfaces) have been carried out by \cite{Hartman:2013qma}\cite{Liu:2013iza}\cite{Leichenauer:2014nxa}.

Codimension one surfaces with $(D-2)-$sphere symmetry are simply geodesics in the metric $ds^2 = -r^{2(D-2)}f(r)dt^2 + r^{2(D-2)}f^{-1}(r)dr^2$. Such curves are described by $r(\lambda),t(\lambda)$, where $\lambda$ is a length parameter. Derivatives with respect to $\lambda$ will be represented by dots. The conserved quantity corresponding to time-translation invariance is $E = r^{2(D-2)}f(r) \dot{t}$, and the parameterization constraint is $r^{2(D-2)}\dot{r}^2 = f(r) + E^2 r^{-2(D-2)}$.

In order to demonstrate (\ref{tocheck}), we consider a maximal surface connecting $(u_L,0)$ to $(0,v_R)$. We will use the boost symmetry to set $u_L = v_R$. The surface connecting the points is characterized by an energy $E$. The volume is
\be
V(E) = 2\int_{r_{turn}(E)}^{r_h}\frac{dr}{\dot{r}} = 2\int_{r_{turn}(E)}^{r_h} \frac{ r^{2(D-2)}dr}{\sqrt{E^2 + r^{2(D-2)}f(r)}}.\label{V(E)}
\ee
where $r_{turn}$ is the turning point at which the denominator vanishes, and $r_h$ is the horizon radius.

To find the volume as a function of $u_L = v_R$, we will compute $u_L(E)$ and compare to $V(E)$. Denoting the order-one value of $u = v$ at $r = r_{turn}$ as $u_{turn}$, we use the definition of the Kruskal coordinates (\ref{defofkruskal}) to obtain
\be
\log u_L(E) = \log u_{turn}(E) + \frac{f'(r_h)}{2}\int_{r_{turn}(E)}^{r_h}dr\frac{\sqrt{E^2 + r^{2(D-2)}f(r)}-E}{f\sqrt{E^2 + r^{2(D-2)}f(r)}}.\label{u(E)}
\ee
The integrands in (\ref{V(E)}), (\ref{u(E)}) are both regular near the upper limit of integration. However, as we increase $E$, the turning point approaches $r_m$, where $f(r)r^{2(D-2)}$ has an extremum. Near this point, both integrals develop logarithmic divergences, representing a long surface running close to $r = r_m$. Since $r_m$ is finite and positive, it is clear that the divergences of the two integrals are proportional to each other. Working out this coefficient of proportionality, recalling $\beta = 4\pi/f'(r_h)$ and our choice of boost frame $u_L = v_R$, one finds (\ref{tocheck}).

\end{document}